# On the nature of wettability of van der Waals heterostructures

Meenakshi Annamalai[a], Kalon Gopinadhan[*a,b], Sang A Han[f,g], Surajit Saha[a,e], Hye Jeong Park[f], Eun Bi Cho[f], Brijesh Kumar[#a], Sang-Woo Kim[*f,g], T Venkatesan[*a,b,c,d,e]

Wetting behaviour of surfaces is believed to be affected by van der Waals (vdW) forces; however, there is no clear demonstration of this. With the isolation of two-dimensional vdW layered materials it is possible to test this hypothesis. In this paper, we report the wetting behaviour of vdW heterostructures which include, chemical vapor deposition (CVD) grown graphene, molybdenum disulfide ($MoS_2$) and tungsten disulfide ($WS_2$) on few layers of hexagon boron nitride (h-BN) and $SiO_2$/Si. Our study clearly shows that while this class of two-dimensional materials are not wetting transparent, there seems to be a significant amount of influence on their wetting properties by the underlying substrate due to dominant vdW forces. Contact angle measurements indicate that graphene and graphene-like layered transitional metal dichalcogenides invariably have intrinsically dispersive surfaces with a dominating London-vdW force-mediated wettability. Electric field controlled wetting studies of $MoS_2$/$WS_2$/$SiO_2$/Si heterostructures were performed and no notable changes to the water contact angle was seen with applied voltage although two orders of magnitude change in resistance was observed. We postulate that the highly dispersive nature of these surfaces arising from the predominant London-vdW forces could be the reason for such observation.

## INTRODUCTION

The isolation of graphene[1] has provided an insight and excitement to explore other two-dimensional (2D) materials that go beyond graphene which include h-BN and transition metal dichalcogenides (TMDCs) such as $MoS_2$, $WS_2$, and $WSe_2$.[2,3] The advent of such materials with intriguing properties has opened many avenues for their use in various applications such as transparent electrodes, sensors, flexible and stretchable transistors, logic circuits, light emitting diodes (LEDs), energy storage and energy conversion devices to name a few.[4-6] These layered materials are held together by vdW forces making it possible to isolate/synthesize atomic layers as well as creating unique hybrids. Although there is significant and continued progress in understanding the electronic, optoelectronic, spintronic and mechanical properties of these materials, interactions of such materials with the surroundings remains less explored.[2,5,7-14] In this study, we explore the wetting properties of vdW based single and hybrid structures which is extremely crucial from both fundamental and application perspective for example in conformal coatings, filtration membranes, energy storage devices, gas barriers, bio-sensors etc.[4,6,9,15-19]

In the past, attempts have been made to understand the surface wettability of graphene. Shin *et al.* performed water contact angle (WCA) measurements on epitaxial graphene grown on SiC substrate and showed that graphene is hydrophobic with a water contact angle of 92° when compared to hydrophilic SiC with WCA of 69°. Shin *et al.* also reported a thickness independent contact angle from the measurements done on mono, bi, multi-layer and highly ordered pyrolytic graphite (HOPG).[20] WCA of graphene transferred from copper and nickel to $SiO_2$ was reported by Kim *et al.* as 93.4° ad 90.4° respectively.[21] Rafiee *et al.* demonstrated that graphene is indeed wetting transparent to copper, gold and silicon and such systems are dominated by vdW forces whereas it is non-transparent to glass due to short range chemical bonding.[22] It was also demonstrated that the contact angle of ~6 layers of graphene and above on copper gets approaches bulk graphite value.[22] The debate on intrinsic water contact angle on graphene came about when the effect of airborne contaminants on the wettability of graphene and graphite were shown by Li *et al.*[23] Contrary to popular belief, it was found that the hydrophobic nature of graphene as reported by other groups was mostly due to hydrocarbons when the samples get exposed to ambient conditions.[23] It was shown that thermal annealing and ultraviolet-$O_3$ treatment was effective in removing airborne contaminants and therefore graphene is intrinsically hydrophilic with WCA of 64.4° on HOPG and 59.6° for 2 to 3 layers of graphene on nickel. With sample ageing and exposure to ambient conditions the WCA showed hydrophobic value >90°.[23] Similar study has been reported by Lai *et al.* where evolution of water wettability of graphene from hydrophilic to hydrophobic was shown.[24] A nanoscopic approach using dynamic force spectroscopy measurements using atomic force microscopy (AFM) has also been demonstrated.[24] The first report on the study of surface energy of pristine monolayer graphene grown on copper substrates and aged graphene with exposure to ambient conditions by contact angle

a. NUSNNI-NanoCore, National University of Singapore (NUS), Singapore 117411
b. Department of Electrical and Computer Engineering, National University of Singapore (NUS), Singapore 117583
c. NUS Graduate School for Integrative Sciences and Engineering (NGS), National University of Singapore (NUS), Singapore 117456
d. Department of Materials Science and Engineering, National University of Singapore (NUS), Singapore 117575
e. Department of Physics, Faculty of Science, National University of Singapore (NUS), Singapore 117551
f. Nanoelectronic Science and Engineering Laboratory, Department of Advanced Materials Science & Engineering, School of Engineering, Sungkyunkwan University, Suwon 440-746, Republic of Korea
g. SKKU Advanced Institute of Nanotechnology (SAINT), Center for Human Interface Nanotechnology (HINT), Sungkyunkwan University, Suwon 440-746, Republic of Korea
# Currently at Centre for Nanoscience and Technology, School of Engineering and Technology, Amity University, Gurgaon 122413, India
† Electronic Supplementary Information (ESI) available: See DOI: 10.1039/x0xx00000x
Email: venky@nus.edu.sg; kimsw1@skku.edu; gopinadhan@iitkalumni.org

measurements and estimation of surface energy through Neumann and Owen, Wendt, Rabel and Kaelble (OWRK) models was reported by Kozbial *et al.* showing that monolayer graphene is mildly polar[25] which is surprising as graphene is supposed to be non-polar. The non-transparency of graphene has also been proved through static contact angle measurements and dynamic force spectroscopy studies using AFM.[25] From all of these studies, it is clear that the surface wettability is greatly influenced by the surroundings. Changing the surrounding is clearly one of the ways to understand the surface wettability better.

Going beyond graphene, there are very few studies on the wetting properties of the other classes of 2D materials like h-BN, TMDC etc. Wetting behavior of $WS_2$ and $MoS_2$ on $SiO_2$/Si substrates grown by CVD has been studied where chalcogen (sulphur) substitution with oxygen resulted in hydrophobic to hydrophilic wettability transition.[26] By varying the growth temperature, Gaur *et al.* have demonstrated a change in the nature of the surface; from hydrophilic to hydrophobic as the temperature is increased clearly suggesting the extrinsic nature of the surface.[27] The surface energy of highly crystalline few-layer thick $MoS_2$ was found to be 46.5 $mJ/m^2$.[27] Recently, vdW heterostructures have been fabricated by stacking different 2D materials and in such structures covalent bonds provide in-plane stability and the vdW forces enable to keep the 2D stack together.[28-30] Most studies have focused on electrical properties and device applications in flexible/transparent electronics, optoelectronics, photonics etc.,[31-34] but to the best of our knowledge, there aren't any reports on the wetting properties of such 2D heterostructures. The surface energy of an overlayer 2D material is influenced by the underlying substrate. That raises fundamental questions: what happens to the surface energy when a 2D material forms heterostructures with other 2D materials?

Motivated by such fundamental questions, we have investigated the wettability of graphene and graphene like layered TMDCs ($MoS_2$ and $WS_2$) as individual and hybrid structures on h-BN and $SiO_2$/Si substrates. More specifically, the nature of surface free energy of the structures and the influence of the underlying substrate on the wettability of the overlayer has been studied. The effect of electric-field gating on the wettability of a hybrid structure has also been presented.

## Experimental

### Contact angle and surface energy measurements

The contact angle and surface energy measurements were carried out using the video based fully automated Data Physics optical contact angle microlitre dosing system (OCA 40 Micro). Solvent drops (1µl/drop) of DI water (more polar), ethylene glycol (intermediate polar & dispersive) and diiodomethane (mostly dispersive) with known surface tensions (refer SI Table 1) were dispensed using a Teflon coated motor driven syringe. The contact angles were measured at 22 °C and 45% RH conditions and a video was recorded (72 frames/second) for every dispensed solvent droplet. Any dynamic changes to the droplet on the surface can be precisely observed through this method. This system allows for estimation of contact angles with 0.1° accuracy. The contact angles with three different test liquids and surface free energies (polar and dispersive) were evaluated using the sessile drop technique and OWRK model respectively (refer SI).

### Atomic force microscopy

The AFM images were obtained using JEOL JSPM 5200 with SiN probe tips of resonance frequency ~300 kHz and force constant ~40 N/m.

### Raman spectroscopy

The Raman spectra of the 2D materials and their heterostructures have been recorded using a JY-Horiba LabRAM HR Evolution Raman spectrometer connected with an air-cooled charge coupled device (CCD) detector. The 514.5 nm line of an Ar-ion laser (Lexel 95-SHG Laser) has been used as an excitation source.

### Growth of $MoS_2$ and $WS_2$

High purity precursors of $(NH_4)_2MoS_4$ (Sigma Aldrich), $(NH_4)_2WS_4$ (Alfa Aesar) was dissolved in dimethyl formamide (DMF) solution and the solution was sonicated for 30 min. Prior to film preparation on a substrate, substrates were cleaned and treated with $O_2$ plasma for 15 min. The prepared solution was then spin-coated on the substrates at 4000 rpm and was then baked on a hot plate at 150 °C for 10 min. To anneal the freshly prepared thin films, films were placed in the heat zone of the quartz tube with a gas flow mixture ($Ar/H_2$) at 1 Torr. This is followed by a two-step annealing process. The chamber was heated up to 400 °C at a heating rate of 5 °C/min and continued for 30 min to efficiently remove the residual solvent, $NH_3$ molecules, and other by-products dissociated from the precursors. After the first annealing step, the gas environment was changed to $Ar/H_2S$ at 1 Torr and the heat zone was heated up to 1000 °C at a heating rate of 20 °C/min and dwelled for 30 min.

### Growth of h-BN

Hexagonal boron nitride (h-BN) nanosheets were synthesized by chemical vapor deposition (CVD) using borazine ($B_3H_6N_3$) as a precursor. The synthesis of the h-BN layer was carried out in a low pressure CVD system. For the growth process, a Cu-foil (Alfa Aesar, 125 µm-thick) was placed in the CVD chamber and gradually heated to 900 °C for 1 h in the presence of Ar (25 sccm) and $H_2$ (25 sccm) gas flow to remove impurities. The temperature was then cooled down to 400 ℃ under the same condition. Once the temperature reached to 400 ℃, 5 sccm borazine was injected with Ar gas flow for 15 min. Subsequently, the temperature was increased to 1000 ℃ with a dwell time of 1 hour. After an hour the chamber was cooled to room temperature.

### Preparation of heterostructures

The single and hybrid structures were prepared by wet transfer method. In this method, poly methyl methacrylate (PMMA) was coated on as-grown $MoS_2$ and $WS_2$ and was baked for 10 min on a hotplate. Then, PMMA/$WS_2$/$SiO_2$/Si and PMMA/$MoS_2$/$SiO_2$/Si substrates were put into BOE (Buffered Oxide Etch) solution for 30 min. PMMA/$MoS_2$ and PMMA/$WS_2$ layers were transferred onto as-grown h-BN/Cu substrate. The sample was dipped in acetone to remove PMMA and the $MoS_2$/h-BN/Cu and $WS_2$/h-BN/Cu heterostructure/substrates were then baked for 15 min subsequently.

## Results and discussion

2D materials and their heterostructures have been transferred on SiO$_2$/Si substrates, as described in the methods section, in order to investigate the nature of interaction between the layers, the surface wettability and surface energy of the over-layers. The quality of the 2D materials and their number of layers has been determined by Raman spectroscopy which is a nondestructive and noncontact technique. Fig. 1 (a-d) shows the Raman spectra of all the fabricated samples used in this study which were transferred on to h-BN and SiO$_2$/Si platforms. The Raman spectrum of graphene [Fig. 1(a)] consists of two unique signatures, namely, G-peak at ~1586 cm$^{-1}$ and 2D-peak at ~2685 cm$^{-1}$.[35, 36] The intensity ratio of the G-peak to the 2D-peak indicates the graphene to be a monolayer.[35, 36] Further, the very weak intensity of the D-peak at ~1350 cm$^{-1}$ is a clear indication of a high quality graphene.[35] On the other hand, the MoS$_2$ and WS$_2$ have unique in-plane ($E_{2g}$) and out-of-plane ($A_{1g}$) vibrations arising from the relative motions of the Mo/W and S atoms. The difference in energy(frequency of vibration) of these two peaks identifies the number of layers being present.[37] The heterostructures of the 2D materials being used in our experiments have a 4-layer thick MoS$_2$ and more than 5-layer thick WS$_2$, as suggested by their Raman spectra [see Fig. 1(b-d)]. However, we could not observe any peak relating to h-BN, possibly due to its very low scattering cross-section. The number of layers of h-BN present in these structures was determined using transmission electron microscopy (TEM) and the obtained TEM image is shown in Fig. S3

Fig. 2(a) is a schematic of vapor-liquid-solid interfaces where Θ, $\gamma_{lv}$, $\gamma_{sv}$, $\gamma_{sl}$, are contact angle, interfacial free energies of liquid/vapor, solid/vapor, solid/liquid respectively. The contact angle is related to interfacial energies through Young's equation[38]

$$\gamma_{lv}\cos\theta = \gamma_{sv} - \gamma_{sl} \qquad (1)$$

Fig. 2(b) shows the measured water contact angles of the various 2D materials (graphene, MoS$_2$, WS$_2$) as well as their heterostructures. The images of the contact angles obtained using the other two test liquids (ethylene glycol and diiodomethane) and the measured contact angle values with error bar is shown in Fig. S4 and Table S2. The standard deviation has been calculated based on the variations in the measured left and right contact angles and also taking in to account the instrumental uncertainty of 0.1°. All the samples used in this study have been thermally annealed at 150 °C to remove any aromatic hydrocarbons and moisture that might be present on the surface before the contact angle measurements. A detailed study of the aromatic hydrocarbon contamination on graphite was recently reported by Martin *et al.* through Kelvin probe microscopy and it was shown that such contaminants desorbs at 50 °C.[39] The change in WCA of graphene on copper upon hydrocarbon contamination measured by ellipsometric phase shift was demonstrated.[25] The contact angle of a sample can be invariably restored and obtained over several runs of measurements after simple thermal annealing to remove any adsorbed molecules on the surface and hence the reproducibility in measurements is ensured [refer to Table S2].

In 2D structures, the in-plane atoms are covalently bonded and the interlayer coupling is through vdW interactions. From Tables 1 & 2, it is clear that these vdW structures are not fully wetting transparent to the underlying substrate as the substrate does influence the surface wettability of these structures. We made sure, that each material (graphene, MoS$_2$ and WS$_2$) on the two different platforms were of same thickness (see Fig. 1) in order to make a comparison on the wetting and henceforth the influence of the underlying substrate on the wettability. Samples deposited on h-BN shows higher wettability when compared to samples of same thickness on SiO$_2$/Si substrates. For example, the hybrid structure of MoS$_2$ and WS$_2$ on h-BN shows a water contact angle of 89.9° degrees whereas this hybrid structure with exactly the same thickness on SiO$_2$/Si shows a water contact angle of 63.3°. This could be attributed to the water contact angle and total surface energy of these underlying systems. This study also provides information about the long range nature of the forces involved in the wettability which essentially allows the substrate surface

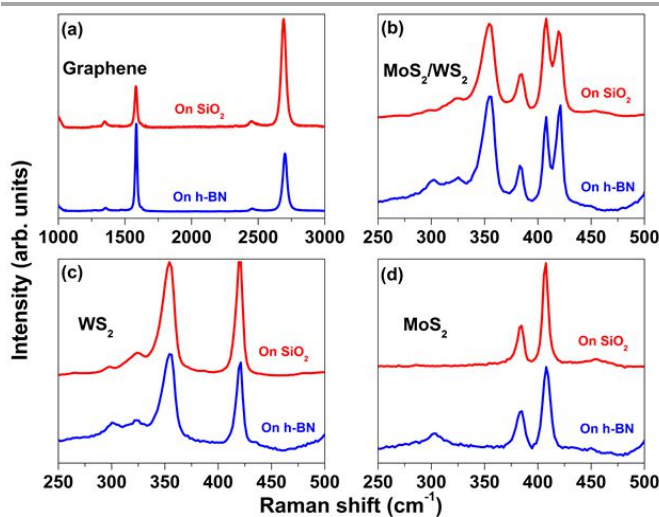

**Fig. 1** Raman spectrum of (a) graphene (either 1 layer or doped bilayer) (b) MoS$_2$ & WS$_2$ (~4 layers of MoS$_2$ and > 5 layers of WS$_2$), (c) WS$_2$ (> 5 layers) (d) MoS$_2$ (~4 layers) transferred on SiO$_2$/Si and h-BN.

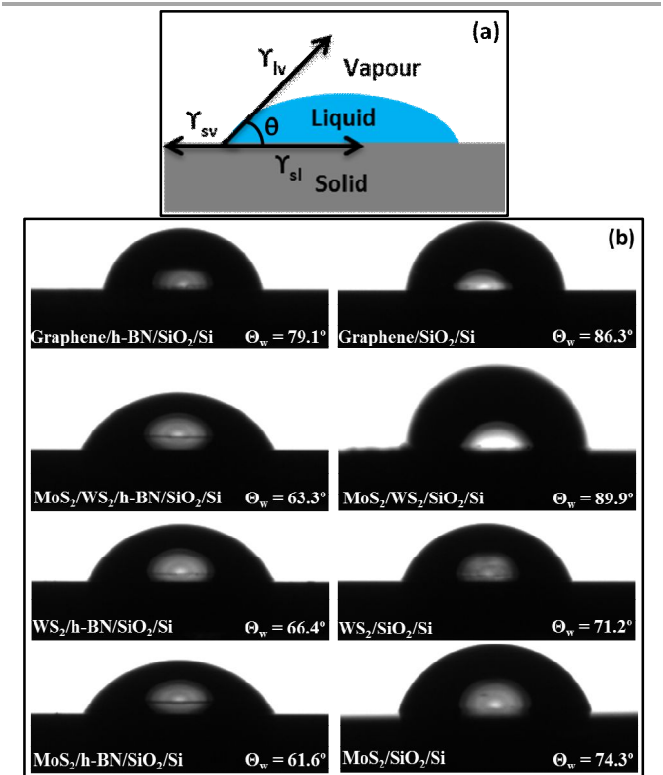

**Fig. 2** (a) Schematic of solid-liquid interface showing the interfacial free energies of liquid/vapor ($\gamma_{lv}$), solid/vapor ($\gamma_{sv}$), solid/liquid ($\gamma_{sl}$) (b) water contact angles obtained on various fabricated structures.

energy to influence 4 to 5 layer $MoS_2$ or $WS_2$ systems. Table 1 summarizes the results of contact angle measurements obtained with three different solvents on $SiO_2$/Si and h-BN/Cu. From Table 1, it can be seen that h-BN is more wettable and has higher total surface energy when compared to $SiO_2$/Si. Kozbial et al. reported that graphene is intrinsically mildly polar although it has non-polar $sp^2$ structure and suggest that the mild polarity could be due to π hydrogen bonding and/or surface defects leading to partial wetting transparency.[25] Our studies show that the surface wettability of both graphene and graphene-like layered TMDCs is predominantly influenced by the wettability of the underlying substrate and this is attributed to its partial wetting transparency.

In order to compute the surface free energy of a solid surface ($γ_{sv}$), the other two energies which include the solid/liquid interface free energy ($γ_{sl}$) and liquid/vapor interface free energy ($γ_{lv}$) in the Young's equation must be known. Clearly, $γ_{lv}$ is known from liquid surface tension. However, $γ_{sl}$ is an unknown variable based on solid surface tension ($γ_{sv}$), liquid surface tension ($γ_{lv}$) and the interactions between two phases. As per Fowkes method, these interactions can be interpreted as the geometric mean of a dispersive and polar component of surface tensions $γ^D$ and $γ^P$ respectively. The solid/liquid interfacial energy can be written as

$$γ_{sl} = γ_{sv} + γ_{lv} - 2\left(\sqrt{γ_{sv}^D γ_{lv}^D} + \sqrt{γ_{sv}^P γ_{lv}^P}\right) \qquad (2)$$

Therefore, two or more liquids with known polar and dispersive components are used to find the solid surface free energy using OWRK model.[40] The OWRK method separates the interfacial surface tension according to the underlying interactions (polar and dispersive) between the molecules. Hence, the total surface energy of the solid is the sum of polar and dispersive components. The polar component originates from the permanent dipole-dipole interactions. They are stronger in polar molecules (having permanent dipole moment). On the other hand, the dispersive component arises due to random fluctuations in the electron density when brought together, leading to an induced dipole-dipole interaction (London-vdW forces or London dispersion forces). The detailed description of OWRK model and fitting results are shown in SI. Importantly, the convergence factor (indicated in Table 2) is very close to ~ 1, showing this model is very suitable to calculate surface energies of vdW structures with a high degree of accuracy.

Fig. 3 shows the surface energy plots of all the fabricated samples fitted using OWRK model and the corresponding surface energy values computed are shown in Table 2. The contact angles obtained on the fabricated 2D structures with three test liquids (water, ethylene glycol and diiodomethane) and the repeatability in water contact angle (water CA repeat) values which were obtained after thermal annealing at 150 °C before every measurement cycle to eliminate the aromatic hydrocarbons and adsorbed water if any is shown in Table S2. The total surface energy values of these structures range from 34 mN/m to 44 mN/m out of which the polar contributions are ~few mN/m. From the surface energy estimations, it is found that all the vdW individual and hybrid structures studied have intrinsically highly dispersive surfaces and this is attributed to the long range London dispersion forces. vdW interactions are an important component of surface forces in these systems. Due to the highly dispersive nature of these surfaces, the classical approach based on Lifshitz and Hamakeron London dispersion forces can be applied and these systems do not require a non-pair wise additive theory where Keesom (force between two permanent dipoles) and Debye (force between a permanent and a corresponding induced dipole) contribution of polar interactions are significant.[41] However, a small increase in the polar component of the surface energy may be noticed when the 2D materials or their heterostructures are placed on h-BN platform. The increase may be due to the polar nature of the underlying h-BN inducing a partial polar character to the overlayer (2D materials/heterostructure). While looking at the structure of these fabricated materials, h-BN, graphene, $MoS_2$ and $WS_2$ invariably possess hexagonal lattice structure.[42-44] While h-BN has highly polar B-N bonds, graphene, $MoS_2$ and $WS_2$ exhibit predominant non-polar bonds. Graphene has homonuclear C-C intralayer bonds and the bonds in such homonuclear diatomic molecules are non-polar.[42-44] $MoS_2$ crystal consists of S-Mo-S sandwiches where S and Mo atoms are held together by covalent bonds.[43] Two surfaces originate from the rupture of vdW interactions (face) and covalent bonds (edge).[43] The face ([0001] plane) generated by the rupture of vdW forces gives the material a non-polar surface. The surface (edge) created by the covalent bonds tend to be hydrophilic.[43] But the wettability in this material is strongly influenced by the vdW interactions, giving it a dispersive nature. $WS_2$ also belongs to the same family of layered TMDCs as $MoS_2$ and consists of S-W-S sandwiches. The edge structures of $WS_2$ are [10$\bar{1}$0] w-edge and [$\bar{1}$010] s-edge.[44] Fig. 4 shows the AFM topography images obtained on all samples used in this study. The wetting characteristics of a solid structure are found to be dependent on the roughness of its surface and such extrinsic influence is inherent to the fundamental wetting phenomena.[45] The root mean square (RMS) roughness of all the fabricated samples are approximately few nanometers or less and therefore the effect of this extrinsic factor on the contact angle of these structures are insignificant.

**Table 1.** Data obtained from wetting measurements with three solvents and calculated surface energies of SiO$_2$/Si and BN/Cu.

| S/N | Sample Details | Water CA (degrees) | Ethylene Glycol CA (degrees) | Diiodomethane CA (degrees) | Surface Energy Total | Surface Energy Polar | Surface Energy Dispersive | Convergence |
|---|---|---|---|---|---|---|---|---|
| 1 | SiO$_2$/Si | 48.4±0.9 | 43.0±0.1 | 49.2±0.9 | 45.04 | 24.47 | 20.56 | 0.9165 |
| 2 | BN/Cu | 21.2±1.2 | 25.9±0.6 | 16.7±0.1 | 58.67 | 30.07 | 28.57 | 0.8616 |

**Table 2.** Calculated surface energies of the fabricated samples using the contact angles measured with three different solvents.

| S/N | Sample Details | Surface Energy Total | Surface Energy Polar | Surface Energy Dispersive | Convergence |
|---|---|---|---|---|---|
| 1 | Graphene/h-BN/SiO$_2$/Si | 37.32 | 3.37 | 33.95 | 0.9881 |
| 2 | MoS$_2$/h-BN/SiO$_2$/Si | 43.17 | 10.56 | 32.61 | 0.9484 |
| 3 | WS$_2$/h-BN/SiO$_2$/Si | 39.98 | 8.09 | 31.89 | 0.9187 |
| 4 | MoS$_2$/WS$_2$/h-BN/SiO$_2$/Si | 43.09 | 8.35 | 34.74 | 0.9111 |
| 5 | Graphene/SiO$_2$/Si | 34.63 | 1.54 | 33.09 | 0.9725 |
| 6 | MoS$_2$/SiO$_2$/Si | 35.35 | 5.79 | 29.57 | 0.9518 |
| 7 | WS$_2$/SiO$_2$/Si | 40.31 | 5.33 | 34.98 | 0.9561 |
| 8 | MoS$_2$/WS$_2$/SiO$_2$/Si | 37.27 | 0.39 | 36.88 | 0.9912 |

The other aspect of this work is to study the wetting characteristics of 2D heterostructures as these materials offer a remarkable electrostatic modification of their properties.[46] For example, the band gaps can be tuned by external electric field and biaxial strain. The band gap change has been attributed to the charge redistributions induced by the external electric field. Field-effect transistors (FETs) based on these 2D materials with TMDC as a channel material, h-BN as gate dielectric and graphene as source, drain and gate contacts have been recently demonstrated and hence making them an interesting alternative material for nanoelectronic applications.[47] Motivated by this, wetting measurements on a hybrid structure comprising of MoS$_2$ and WS$_2$ on SiO$_2$/Si platform was performed with applied gate voltage and the data obtained is shown in Fig. 5 for a measurement span of 176 seconds in the first run [Fig. 5 (b)] and for a much lesser time frame in following runs (run 2 ~36 seconds and run 3 ~15 seconds)[Fig. 5(c-d)]. The measurements were repeated for three different cycles and the WCA was monitored with applied voltage. Apparently the change in contact angle was observed due to the rate of evaporation of the water and was not dependent on the applied voltage although the change in current was two orders of magnitude with applied voltage [Fig. 5(a)]. The rate of change was contingent on the measurement duration and for a longer measurement span the WCA changed from 89.9° to 55.5° whereas for short duration WCA change was observed to be only few degrees. Before starting every cycle of measurement, the samples were thermally annealed under similar conditions as mentioned previously and the initial WCA obtained at 0 V for all three runs is clearly reproducible [Fig. 5(b-d)]. We postulate that the highly dispersive nature of the surfaces due to dominating London–vdW forces could be the cause of such observation as no electric field effects are expected.

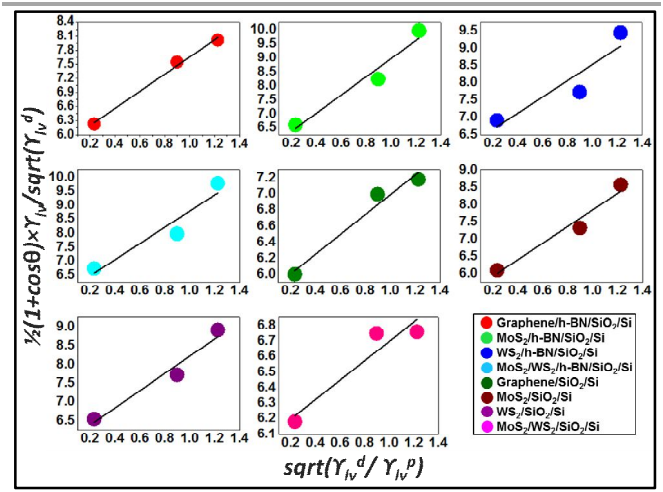

**Fig. 3** Surface energy (polar and dispersive) graphs obtained on various samples based on OWRK model. Convergence values shown in Table 2 indicate the accuracy of the model.

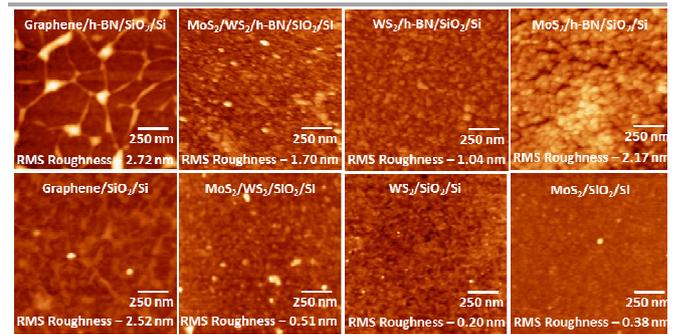

**Fig. 4** AFM topographic images of the various fabricated surfaces.

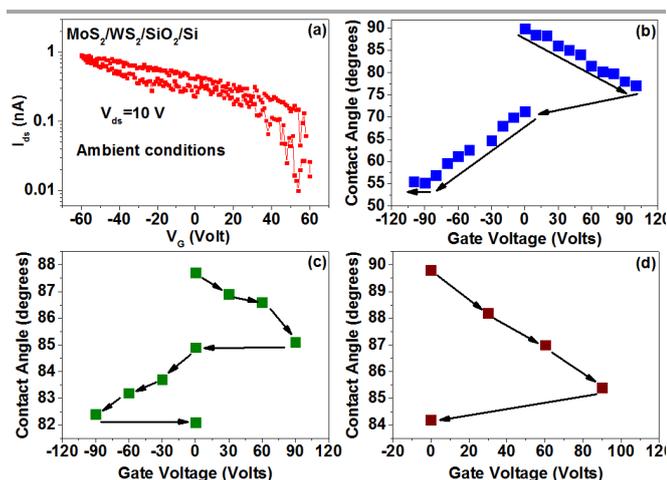

**Fig. 5.** Data obtained from a $MoS_2/WS_2/SiO_2/Si$ heterostructure (a) The gate voltage ($V_G$) - drain current (Ids) characteristics (b- d) water contact angle with applied gate voltage from three measurement runs.

## Conclusions

We have explored the wettability of vdW based individual and hybrid structures. Our measurements clearly show that the 2D materials are not totally wetting transparent but there is measurable influence of the underlying substrate on the wettability making them partially wetting transparent. From surface energy calculations it is found that invariably the major class of 2D materials which include graphene, $MoS_2$, $WS_2$ possess dispersive surfaces and hence we propose that the lack of change in wettability by electric-field gating is attributed to their dispersive nature. However, these 2D materials when placed on h-BN, reveal a non-negligible polar component perhaps induced by the polar nature of the underlying h-BN. We believe our work will help the scientific community to further understand the fundamentals of liquid-solid interactions in the exciting class of 2D materials and pave the way for future applications involving solid/liquid interfaces.

## Acknowledgements

We would like to thank Centre of Integrated Circuit and Failure Analysis (CICFAR) for providing the AFM facility. M. Annamalai would like to acknowledge support from Nanoscience & Nanotechnology Institute (NUSNNI) core fund. T. Venkatesan would like to acknowledge the funding support from National Research Foundation Competitive Research Programme (NRF-CRP8-2011-06) and Singapore-Berkeley Research Initiative for Sustainable Energy (SinBeRISE) NRF CREATE programme.

# Supplementary Information

## Macroscale contact angle and surface energy measurement

Wettability of surfaces by liquids is of great interest in a number of fields ranging from engineering to medicine. Wetting phenomena on a macroscopic scale can be illustrated using Young's equation.[48]

$$\gamma_{lv}\cos\theta = \gamma_{sv} - \gamma_{sl} \qquad (1)$$

where θ is the contact angle, $\gamma_{sl}$ is solid/liquid interfacial free energy, $\gamma_{lv}$ is liquid/vapourvapourinterfacial tension (surface tension) and $\gamma_{sv}$ solid surface free energy.

The contact angle is estimated using the sessile drop technique by measuring the angle between the tangent lines along solid-liquid interface and liquid-vapour interface of the liquid contour as shown in FigureS1. A contact angle of 0º and 180º correspond to complete wetting and non-wetting respectively. Surfaces exhibiting contact angles below 90º are called hydrophilic and those above 90º are called hydrophobic. In the past few decades many semi-empirical analytical models have been developed to compute surface free energy from measured contact angles such as Fowkes,[49] Owens Wendt Rabel and Kaelbel (OWRK),[40] Van Oss-Chaudhury Good/Lewis acid base theory,[50] Zisman[51] and Neumann,[52] to name a few. Each approach is targeted for measuring surface energies of either low surface energy materials or high surface materials or both. Also, the essence and physical interpretation of these approaches are different and therefore subsequently provide information on total surface energy or individual components (polar, dispersive, hydrogen bonding etc.) of surface energy or both. In our work, the OWRK method has been adopted which is suitable for universal systems.

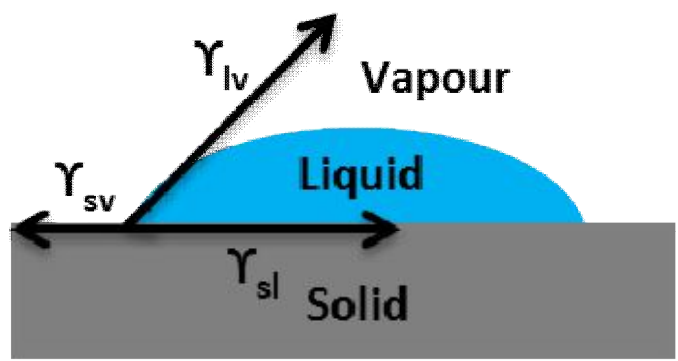

**Figure S1.** Schematic diagram of a liquid drop on a solid surface showing the interfacial tensions at three phase boundary.

## Description of OWRK method

Owens, Wendt, Rabel and Kaelble developed a two component model to separate the interfacial tension according to the underlying interactions between the molecules. These interactions are defined as polar and dispersive interactions. The total surface energy of the solid is the sum of the two parts. The polar interactions arise due to the permanent dipole – permanent dipole interactions or Keesom forces.[53] They are stronger and only exist in polar molecules. Dispersive component also known as London forces are weak and arise due to random fluctuations in the electron density in an electron cloud and hence lead to temporary/induced dipole interactions.[54]

In OWRK method, at least two liquids with known dispersive and polar parts of surface tensions are needed to compute the solid surface free energy as there are two unknowns (solid/liquid interfacial free energy and solid surface free energy). The combining rule proposed by OWRK model is indicated below.

$$\gamma_{sl} = \gamma_{sv} + \gamma_{lv} - 2\left(\sqrt{\gamma_{sv}^D \gamma_{lv}^D} + \sqrt{\gamma_{sv}^P \gamma_{lv}^P}\right) \quad (2)$$

Where $\gamma_{sv}^D$ and $\gamma_{lv}^D$ are dispersive components and $\gamma_{sv}^P$ and $\gamma_{lv}^P$ are polar components of solid and liquid surface energies respectively.

Substituting for $\gamma_{sl}$ from equation (1),

$$\sqrt{\gamma_{sv}^D \gamma_{lv}^D} + \sqrt{\gamma_{sv}^P \gamma_{lv}^P} = \frac{1}{2}[\gamma_{sv} + \gamma_{lv} - (\gamma_{sv} - \gamma_{lv}\cos\theta)] \quad (3)$$

$$\sqrt{\gamma_{sv}^D \gamma_{lv}^D} + \sqrt{\gamma_{sv}^P \gamma_{lv}^P} = \frac{1}{2}[\gamma_{lv}(1 + \cos\theta)] \quad (4)$$

By dividing $\sqrt{\gamma_{lv}^D}$ in equation (4), we get,

$$\sqrt{\gamma_{sv}^D} + \sqrt{\gamma_{sv}^P}\sqrt{\frac{\gamma_{lv}^P}{\gamma_{lv}^D}} = \frac{1}{2}\frac{[\gamma_{lv}(1+\cos\theta)]}{\sqrt{\gamma_{lv}^D}} \quad (5)$$

The above equation can be represented in the linear form,

$$c + mx = y \quad (6)$$

where in, $c = \sqrt{\gamma_{sv}^D}$ $m = \sqrt{\gamma_{sv}^P}$ $x = \sqrt{\frac{\gamma_{lv}^P}{\gamma_{lv}^D}}$

A graphical representation of the OWRK method is shown in FigureS2(MoS$_2$ on SiO$_2$-Si). The polar and dispersive components of total surface tension of the liquids used in this study are known (Table 1) and are substituted to compute the polar and dispersive components of the surface free energy of the solid. The slope of the graph gives the polar component and the vertical intercept gives the dispersive component of the solid surface free energy.

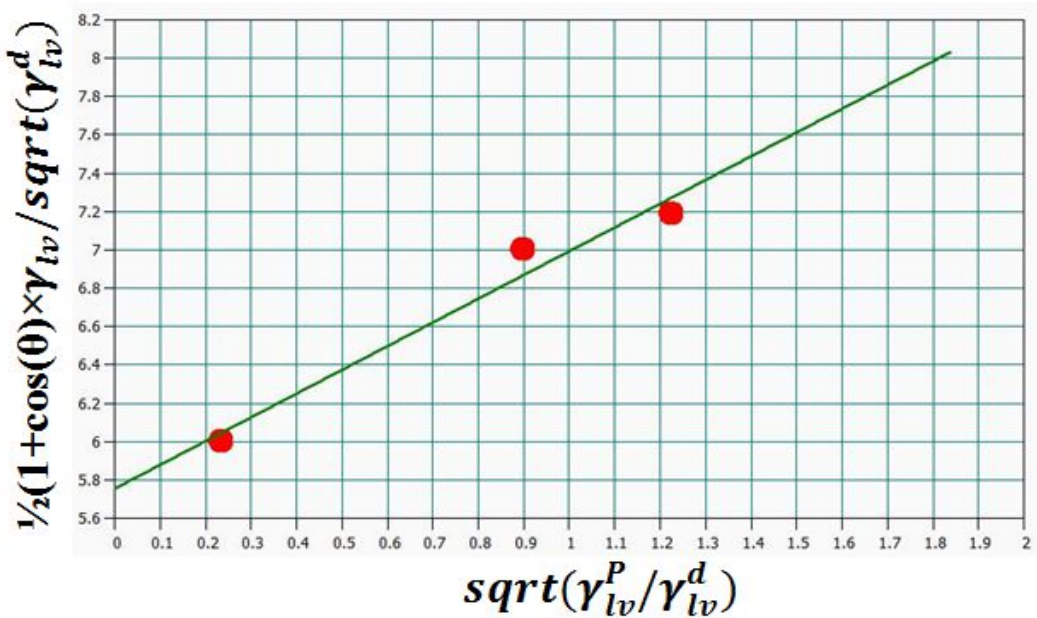

**Figure S2**. A representative surface energy (polar and dispersive) calculation graph based on OWRK model for MoS$_2$/SiO$_2$/Si sample.

**Table S1.** Surface tensions (SFT) of the test liquids

| Solvent | SFT (Total) mN/m | SFT (Dispersive) mN/m | SFT (Polar) mN/m | Values Adapted From |
|---|---|---|---|---|
| Water | 72.80 | 29.10 | 43.70 | Chen et al. |
| Ethylene Glycol | 47.70 | 26.40 | 21.30 | Gebhardt et al. |
| Diiodomethane | 50.00 | 47.40 | 2.60 | Busscher et al. |

For determining the thickness of h-BN, h-BN has been transferred to a copper grid for TEM characterization. The high resolution TEM image on the edge of h-BN is shown Figure S3. The image shows the crystalline and layered nature of h-BN where the number of BN layers is determined to be ~10.

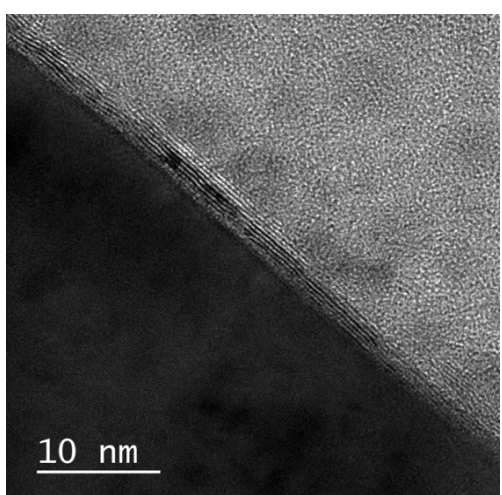

**Figure S3.** A transmission electron microscope (TEM) cross sectional image of h-BN.

Measured contact angles with three test liquids on the fabricated 2D structures are shown in Table 2. The standard deviation has been calculated based on the variations in the measured left and right contact angles and also taking in to account the instrumental uncertainty of 0.1º. Figure S4 shows the contact angle images obtained for ethylene glycol and diiodomethane on the 2D structures.

**Table S2.** Contact angles obtained from three different solvents on the fabricated surfaces

| S/N | Sample Details | Water CA (degrees) | Water CARepeat (degrees) | Ethylene Glycol CA (degrees) | Diiodomethane CA (degrees) |
|---|---|---|---|---|---|
| 1 | Graphene/h-BN/SiO$_2$/Si | 79.1±0.2 | 78.4±0.4 | 51.1±0.6 | 44.3±0.5 |
| 2 | MoS$_2$/h-BN/SiO$_2$/Si | 61.6±0.2 | 60.6±0.2 | 39.3±0.2 | 35.1±0.3 |
| 3 | WS$_2$/h-BN/SiO$_2$/Si | 66.4±0.4 | 68.5±0.2 | 47.7±0.5 | 38.6±0.4 |
| 4 | MoS$_2$/WS$_2$/h-BN/SiO$_2$/Si | 63.3±0.4 | 60.7±0.2 | 44.1±0.2 | 31.1±0.1 |
| 5 | Graphene/SiO$_2$/Si | 86.3±0.2 | 84.6±0.8 | 59.5±0.9 | 49.3±0.2 |
| 6 | MoS$_2$/SiO$_2$/Si | 74.3±0.1 | 76.6±1.1 | 54.8±0.3 | 47.3±1.4 |
| 7 | WS$_2$/SiO$_2$/Si | 71.2±0.2 | 68.5±0.2 | 48.4±1 | 36.7±0.1 |
| 8 | MoS$_2$/WS$_2$/SiO$_2$/Si | 89.9±0.5 | 87.7±0.4 | 63.0±1 | 45.4±0.6 |

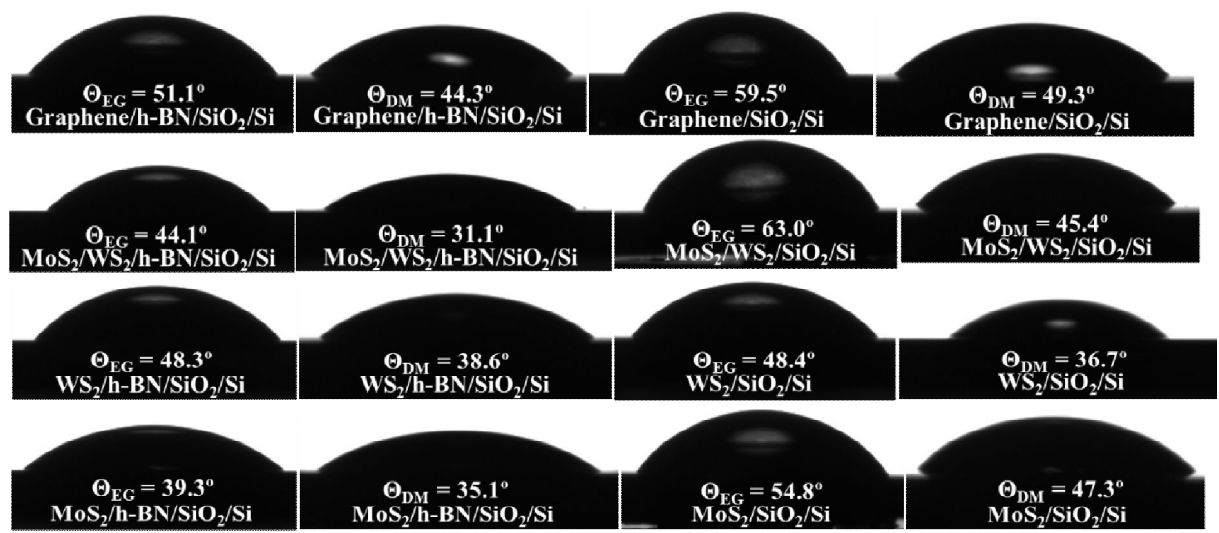

**Figure S4.** Contact angles obtained on various fabricated structures with ethylene glycol and diiodomethane.